# THEORY OF SUPERSOLIDITY

Philip.W.Anderson,.Princeton.University.

Abstract
After reviewing some experimental facts, and early theories, I  sketch the Hartree-Fock description of Boson solids, emphasizing the contrast with the Fermion case in that the natural solution is a product of local wave-functions.  I then investigate the Boson Hubbard model to demonstrate that the local functions are not orthogonal and to exemplify a novel state of matter which has superflow but not ODLRO, below a cooperative thermal transition.  I then discuss whether or not these ideas actually apply to solid He.

## Some Experimental Facts

"Supersolidity" , according to our present understanding, may be generic in Bose solids, but it is experimentally available only in solid He-4, and even there it is extremely sensitive to He-3 impurity level and to defect density in the crystal.  The experimental signature is temperature and amplitude dependence of the moment of inertia as measured, usually, in a highly sensitive torsional oscillator containing an annular sample of solid He.  The characteristic behavior of the moment of inertia as a function of temperature and amplitude of oscillation is shown in figs 1 and 2, the first from the 2004 discovery paper by Kim and Chan[1], the second from a recent study in J C Davis' lab. [2] Dissipation accompanies the decrease in moment of inertia; this is also shown in the figures.  My reading of the experiments is that the dissipation is markedly asymmetric as a function of T and amplitude, suggesting that it is mainly confined to above a transition, at higher T and amplitude, and that there is then  a true phase transition near 50 mK into a  broken symmetry state with dissipationless transport;  this interpretation is supported by the observation of hysteresis below about 50mK.[3]

From the first, I have assumed that the observations at temperatures up to 200 mK and higher do not indicate a phase transition, but rather an extended range of dissipative nonlinear behavior, similar to that observed by Ong 's group in cuprate superconductivity[4], and there described as "vortex liquid" phenomenology.  The Tc for a true phase transition, if it exists, is in the range 50 mK.

Some further observations that seem significant:   first, the early experiments followed very similar observations on solid He in porous media,[5] where any dislocation motion would be unlikely.  Second, recent observations by Kubota's lab and Kim et al[6] show a sensitivity to DC rotation which seems to clinch that it is a quantum vorticity effect.  But , third, Beamish et al[7] have shown that there is low temperature stiffening of the shear elasticity which seems to track the TO effects reasonably well, and is attributed by them to dislocation pinning by He-3 impurities.

Finally, there is a strong dependence on He-3 concentration. Commercial He-4 contains about $.3 \times 10^{-6}$ He-3, and exhibits about 1% decrease in moment of inertia  as T->0 (NCRI).  If we reduce the concentration by a factor 300, the NCRI reduces to about .04%, and seems to be characteristic of the pure crystal.

Theoretical Considerations

Early papers by Reatto and Chester, based on a Jastrow-Bijl approximation to the ground state wave function, proposed the existence of off-diagonal long-range order (ODLRO) even in the perfect crystal (papers by Andreev had suggested already in 1967 that finite concentrations of vacancy defects would superflow.)  But it has never been demonstrated that  ODLRO, superfluidity, and NCRI are synonymous, and the indirect nature of these arguments leaves one, at best, unsatisfied.

The development of simulation methods over the years has led to very good agreement on the physical properties of the solid per se using reasonable potentials of interaction. The pure Bijl-Jastrow-MacMillan wave functions do not quite give perfect agreement and it seems necessary for the best results to, in one way or another, include a single-particle periodic structure a priori. Many of the simulators assert quite strongly that they find no evidence for superfluidity, but it is legitimate to question whether the $10^{-4}$ magnitude of the effect and the proposed temperature scale of $\approx 50$mK are yet achievable, and none of these papers has attacked the problems of principle involved. Some of them in fact seem irrelevant , since they ignore the possibility of a cooperative thermal phase transition like the magnetic transition in He-3.[8]

Since about 1959 it has been clear that superfluidity and superconductivity are caused by macroscopic quantum coherence, and that the marker for these states is the existence of a stiffness against phase rotation of the relevant conserved quantum field, in the case of He4 the Bose field of the particles. A representation of the wave function which allows us to investigate that question is the Hartree-Fock scheme for bosons which was sketched in my book's Ch IV[9] a number of years ago. In that scheme one assumes that the ground state is a product of bosons, one for each site on a lattice, and derives wave equations for the occupied site states as Hartree-Fock equations for holes. Unlike Hartree-Fock theory for Fermions, the different site functions satisfy completely different equations, and these differ from the equations for particles, which have the symmetry of the lattice. Therefore there is a large energy gap even within the mean field theory: the hole excitations can be localized in self-consistently created potential wells, the particle excitations are necessarily free running waves. As was demonstrated in the reference, Goldstone's theorem can be satisfied by an independent spectrum of phonon modes, not necessarily involving the same single-particle excitations.

Fitting equations to the above words, the assumed mean field approximation to the ground state is

$$\Psi_0 = \prod_i b^*_i \, | vacuum \rangle \qquad [1]$$

where

$$b^*_j = \int dr \Psi^*(r) \varphi(r - r_j) \quad where \quad r_j \text{ are the sites}$$

of a regular lattice, and $\Psi$ is the boson field. $\qquad [2]$

The $\varphi(r)$ are to be determined self-consistently
from the solution of the equations of motion for holes:

$$E_i \varphi(r - r_i) = -\hbar^2 (\nabla^2 \varphi_i) / 2m + \overline{V}(r) \varphi_i + \int \overline{A}(r - r') \varphi_i(r') dr' - \Delta V_i \varphi_i -$$

$$\int \Delta A_i (r - r') \varphi_i(r') dr' \qquad [3]$$

Here $\overline{V}$ and $\overline{A}$ are the potential and the exchange operator using the full density summed over all sites j, and $\Delta V_i$ and $\Delta A_i$ are the contributions to th of the boson at site i. V and A are defined simply as

$$V = \int dr' v(r - r') \rho(r')$$

$$A = v(r - r') \rho(r - r'), \qquad [4]$$

v being the potential of interaction and $\rho(r-r')$

the density matrix.

The method is explained in more detail (including normalization, which is a bit tricky) in the reference. There are several crucial points to be noticed. As in the Fermion case, the direct and exchange terms in the self-energy due to the local particle are identical in form, but unlike the Fermion case, where they cancel, they **add** rather than subtract. Thus the potential well for localizing the bosons is twice as strong as it would be in the absence of statistics. In particular, the potential well formed for a substitutional He-3 atom is much shallower than that for an He-4. It is conceivable that the substitutional He-3 atom is not localized at all. I am not aware of any calculations bearing on this question which include this statistical effect. If it is not localized, it cannot serve as a pinning center for dislocations of the parent lattice.

The important thing to realize is that the localized eigenbosons are not automatically orthogonal, in fact they are necessarily NOT orthogonal since each is the lowest eigenfunction of its local potential and therefore is everywhere positive. I will show, by working out a simple soluble case, that therefore the energy necessarily depends on the relative phase of the boson field which thereby becomes a physically relevant degree of freedom of the system and, inter alia, also implies that the ground state is supersolid. My approach to the supersolidity question is often characterized [10] as the introduction of vortices, but the real essence of it is the realization that the local phase is a physical variable, whose low-energy excitations turn out to be vortices, with a finite energy gap due to the core energy.

The Bose-Hubbard Model
The "simple soluble" case which I have studied[11] is that of the Bose-Hubbard model. In this model one assumes that the localizing potentials which give us a periodic lattice of boson wave functions are imposed from some outside source—say an optical lattice in a system of ultracold bosonic atoms. The lowest eigenstates of such a lattice form a band from which we can form localized Wannier functions $\varphi_i(r)$ and thence create a set of orthonormal bosons

$$b_i^* = \int \varphi_i(r)\Psi^*(r)dr \quad with (b_i^*, b_j) = \delta_{i,j} \quad [5]$$

, Then the Bose-Hubbard Hamiltonian is

$$H = \sum_{i,j} t_{ij} b_i^* b_j + U\sum_i \lceil n_i(n_i - 1) + (\varepsilon_0 - \mu)n_i \rceil \quad [6]$$

where we have introduced a strong repulsion U between bosons on the same site, as in the Fermion Hubbard model. In the case where t is very small compared to U, a suitable approximate ground state is the simple product function

$$\Psi = \prod_i b_i^* |vacuum\rangle \quad [7]$$

But [7] is manifestly not the true ground state because it has a finite matrix element $t_{ij}$ to any state in which site i is empty and a neighbor j is doubly occupied. This corresponds to the fact, emphasized in reference 4, that the true ground state of a many-boson system cannot change sign, and the orthogonalized Wannier functions of any simple band must do so. In order to arrive at a better approximation than [7] to the true ground state we must transform to non-orthogonal but more localized site functions $\varphi_i'$. Within the confines of the Hubbard model , this may be done by transforming to a product of **non**orthogonal bosons, where the new bosons are given by

$$b'_i = N^{-1}\lceil b_i + \sum_j t_{ij} b_j / U \rceil; \quad N = \sqrt{1 + \sum \frac{t^2}{U^2}} \quad [8],$$

and the energy gained is

$$\Delta E = -\sum_j t_{ij}^2 / U \qquad [9]$$

By the virial theorem, the gain in kinetic energy is -2 times the loss in potential energy.

The Wannier function, of course, is an equal superposition of every momentum state in the band and therefore has exactly zero net hopping (kinetic) energy. If one were, then, to submit the state [7] to a vector potential A, i e to a net shift of the momentum of every state in the band by A, there would be no change whatever in the net kinetic energy. This state is a true "insulator," a classical solid in the sense of Kohn.[12] The energy may be shown to be unaffected by the relative phasing of the localized, orthogonal bosons; they do not interfere with each other..

The nonorthogonal boson state however is **not** an equal superposition but is weighted toward the bottom of the band; it gains energy by the bosons being equally phased. Calculating the dependence of the energy on the phase angle, it is essential that we realise that we are *not* solving the problem of an isolated pair of sites, but of a pair which we embed in an infinite or toroidally connected sample; we are giving the entire momentum spectrum a boost. Simplest is to consider the problem of a single site and its neighbors along the phase gradient A—which in a simple lattice will come in pairs on opposite sides of the 0 site. The boson belonging to such a site may be simplified to

$b_0' = ub_0 + v[b_1 + b_{-1}]$, and expanding in terms of kinetic energy eigenbosons,

$$b_0' \sim N^{-1/2} \sum_k [u + 2v\cos k]b_k$$

$$E_K = -t/N \sum_k \cos k [u^2 + 4uv\cos k + 4v^2 \cos^2 k]$$

$$= -2tuv$$

[10]

the last equality holding for the ground state. But now if we introduce the momentum shift k->k+A in the wave function the cos² term is multiplied by cos A and the energy [6] likewise: the energy becomes phase-dependent and therefore a vector potential induces a current—obeying, essentially, a London equation J∝A.

I remark on the physical meaning of the extra degrees of dynamical freedom that are implied by the existence of a coupling energy for the phase. The situation with respect to the Fermion and Boson solids has always seemed to me unsatisfactory. He3 has spin degrees of freedom and these undoubtedly order due to particle exchange at temperatures in the millidegree range for the bcc crystal. But until now there has been no corresponding exchange effect for the Bose solid. This seemed to me to be an anomalous asymmetry. The derivation of the coupling effect for the Bose-Hubbard has a close similarity to the derivation of "kinetic exchange" (superexchange)in the Mott insulator.

In the Hubbard model longitudinal currents are strongly forbidden by the interaction "U" which makes the states with integer filling stable and incompressible at almost all values of the chemical potential. But currents can flow in toroidal samples, or vortices can occur if line singularities of the phase field are allowed, which is the required response if the lattice is rotated or sheared.

For definiteness let us model the phase-dependent energy with an x-y model

$$H_{phase} = \sum_{i,j} J_{ij} \cos(\varphi_i - \varphi_j) \qquad [11]$$

which, in 3D and very likely in 2D, has a critical temperature of the order of J , below which the phases are aligned and above which vortex loops proliferate and the flow of current becomes dissipative.

The Phase Variable in Solid He

To what extent can the Bose-Hubbard Model represent the true solid?  It seems reasonable that the self-consistent mean field  orbitals   $\varphi$ (r-$r_i$) can define the local bosons and that they may be improved upon perturbatively.  But the real lattice has phonons as low-energy excitations also  and the question is whether the phonons couple to the vortex excitations of the phase field.  In the fluid they do and the result is a single branch of excitations  with partly  particle and partly phonon character.  I believe, but do not consider it proven, that in the solid the phonons are decoupled from the vortex (roton) excitations.  The longitudinal phonons would couple to the nonexistent compressibility of the vortex liquid, while the transverse phonons of sufficiently low energy have very long wavelength and their effect simply may be represented as a coupling of the phase degree of freedom to shear (see below).

On the other hand, the response to static elastic strain is manifest in the phase Hamiltonian [11].  This part of the Hamiltonian will change by an amount equal to several J per site from temperatures well above Tc  to below it. The coefficients J will be rather steeply dependent on interatomic distances, as is well-known for He3 exchange integrals, which seem to fall off with a very high power—about 15-20—of the distance between sites.   We then realize that the contribution to the stiffness is proportional to the **second derivative** of the exchange constant with respect to the atomic distances, so that the stiffness effect will be at least an order of magnitude—perhaps two—greater than estimates based on the magnitude of NCIR. Another way of seeing this is to realize that the repulsive part of the force between sites comes from the overlap of the two wave-functions on the separate sites.  The exchange contribution to these energies is not necessarily a small correction, as it is not to the depth of the localizing well.

I therefore believe that it is reasonable to conjecture that the low-temperature change in stiffness is not necessarily a structure-sensitive dislocation phenomenon, but may come wholly or in part from the same quantum coherence effect that is manifest in the NCIR.  Recent measurements by Davis et al[13] which show parallels in thermal behavior between the two sets of phenomena, then, need not be interpreted as evidence against quantum coherence vis a vis dislocation pinning effects. I see it as almost impossible to explain the NCRI observations as a dislocation phenomenon, so to me this is a welcome reconciliation.

One difference between this hypothecated state and a superfluid is that the currents would be blocked by any crack as wide as 1 nm, since it requires a continuum of overlapping atoms.  This could be responsible for many of the anomalous experimental results.

Sensitivity to He3 at Extremely Dilute Concentrations.

I consider this problem of sensitivity to be the most profound puzzle in the whole field of supersolidity.  The majority of authors have, for want of other ideas, ascribed it to the effect of pinning of dislocations by fixed He3 "impurities", which has the multiplicative effect of controlling the motion of long lengths of dislocation with few pinning centers.  But the few scenarios for explanation of the phenomenology of the effects which I have seen leave me puzzled as to both the sign and the magnitude of the observed effects.

I would like here to advance a totally different alternative scenario, without claiming that I can give it any  quantitative theoretical support—since all of the quantities on which it relies are certainly uncomputed if not uncomputable.  There are experiments which can differentiate this scenario from others, notably using NMR on the "impurity" He3, and I would like to strongly urge that they be carried out, and to remark that with modern equipment they are by no means out of reach because of the low concentrations involved.

The question this scenario stems from is:  What is the state of He3 atoms in solid He4?  How much evidence is  there that they do indeed form trapped, bound substitutional centers, which can in turn be pinning centers for dislocations.?  As I remarked above, the "Mott" repulsion  parameter U for He4 gains half of its strength from the exchange self-energy, and this half  is absent for a He3 impurity.  The lighter mass would contribute something to delocalization; and solid He is close to the edge of stability. He3 is, as a matter of course, not a static object, it will have some hopping amplitude and a corresponding  quantum  kinetic energy, but the question I am asking is more quantitative:  is it more free-particle-like? He3 dissolves as a Fermi fluid  in superfluid He4 up to 6 %; does it possibly dissolve in supersolid He4?  And in particular, does it dissolve more easily in the phase-ordered state than in the higher-temperature classical solid?  We  can imagine that mobile He3 would actually stir the surrounding He4 and increase the effective exchange amplitude.  The problem, of course, is to explain the multiplicative effect: how does one He3 atom affect the motion of 10,000 He4's?  But it is  not much less plausible than the dislocation effect, in that a small amount of He-3  could lower the free energy of the supersolid, phase-correlated phase sufficiently to trigger the transition into  that phase.  The whole situation is further complicated, as was pointed out to me by Balibar[14] , by the fact that these temperatures are of the same order as phase separation temperatures, so we have no idea of the correct phase diagram in this region of concentration and temperature.

There has been a small amount of NMR work on He-3 in solid He-4, but at concentrations considerably higher than that of natural He-4[15]; even this shows a transition ascribed to phase separation.   But with the many-Tesla magnetic fields now available and at 10-100 mdeg temperature, there should be no signal-to-noise problems  for $10^{-6}$ concentrations.

Conclusions
The scope of this paper ranges from the quite solidly affirmed conclusion about superfluidity of an unfamiliar type in the Bose-Hubbard model, through the quite reasonable conjecture that a renormalized Hartree-Fock  treatment of the real He4 solid would exhibit the same behavior as the simplified Bose-Hubbard model, to admittedly speculative suggestions about the behavior of He3 impurities  and the reasons for the surprising sensitivity of NCRI and shear elasticity to such impurities.  I suggest an experimental investigation through high-sensitivity NMR of the actual state of such impurities.

Acknowledgements
I acknowledge the discussions mentioned in the text with S Balibar,  extensive discussions of their data with Moses Chan,

John Beamish,  Minoru Kubota, and E Pratt, and discussions with B Clark.

.

# Bulk solid helium in annulus

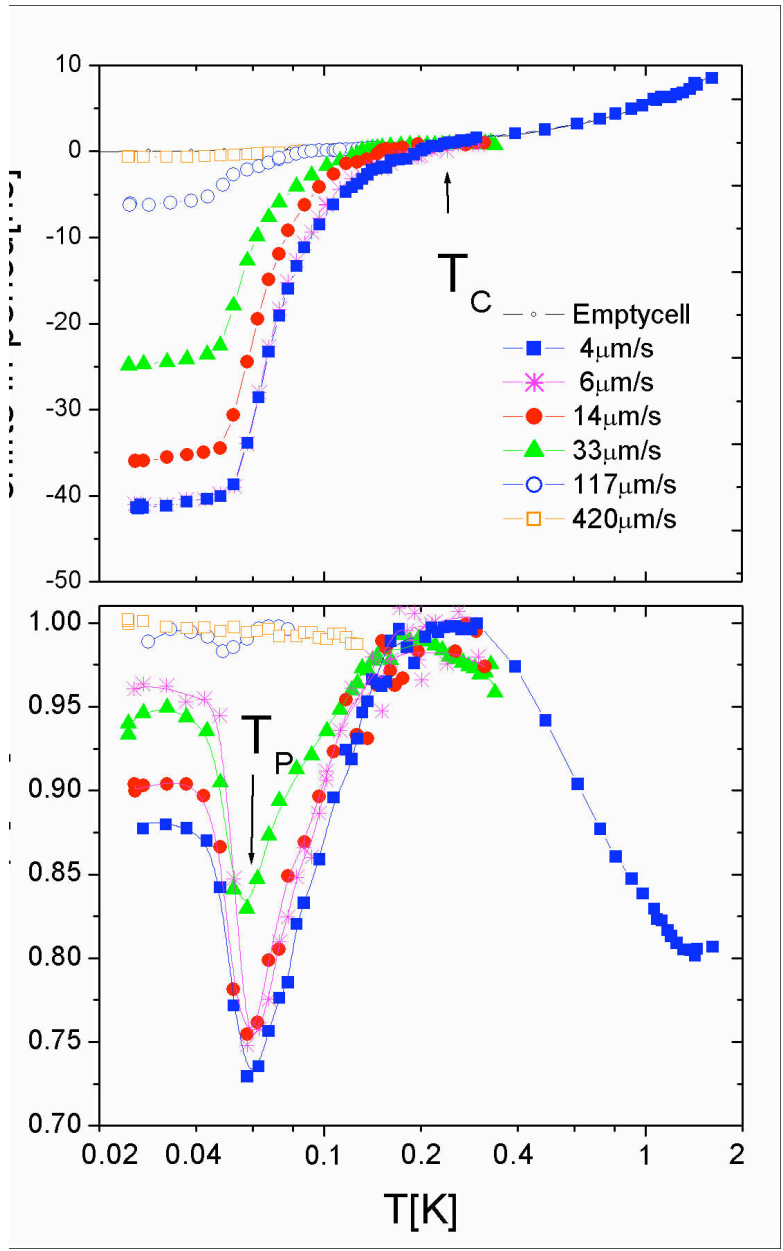

$f_0$=912Hz

51bar
Total mass loading = 3012
Measured decoupling, -Δτ
NCRIF = 1.4%

E. Kim & M.H.W. Chan, Science

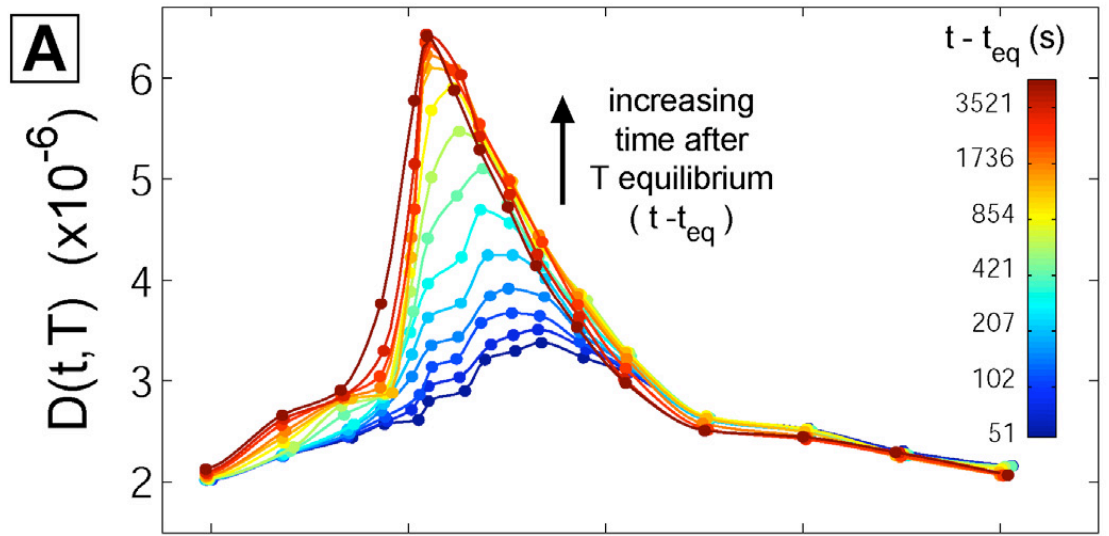

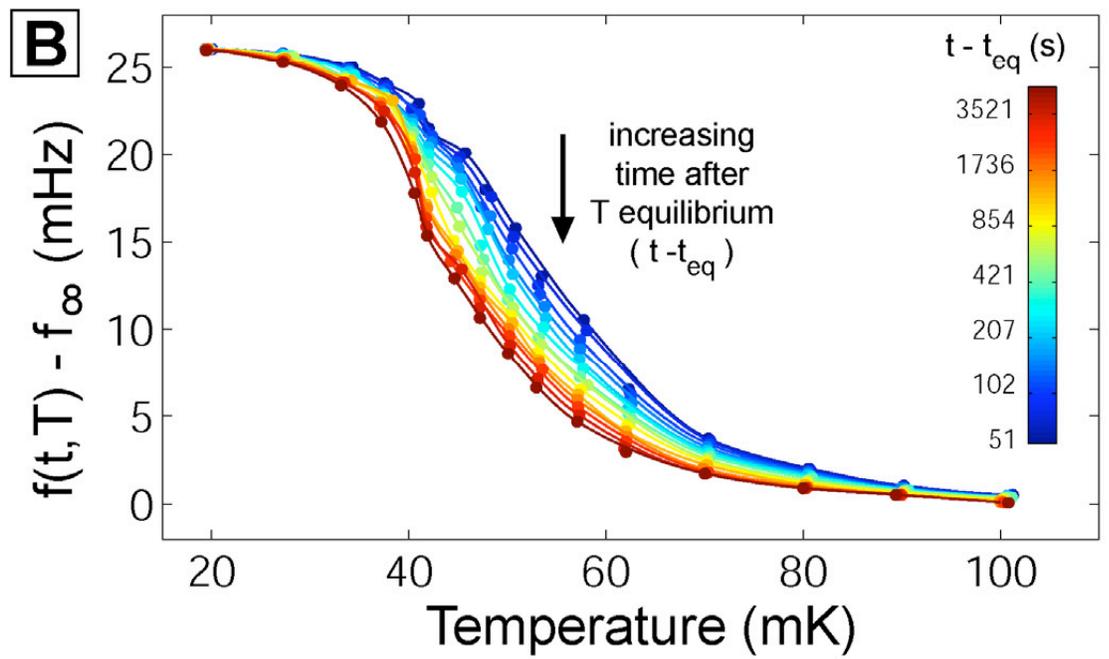